\documentclass[%
 aip,
 jmp,
 amsmath,amssymb,
 reprint,%
]{revtex4-1}

\usepackage{graphicx}% Include figure files
\usepackage{dcolumn}% Align table columns on decimal point
\usepackage{bm}% bold math
\usepackage[utf8]{inputenc}
\usepackage[T1]{fontenc}
\usepackage{mathptmx}
\usepackage{etoolbox}

\makeatletter
\def\@email#1#2{%
 \endgroup
 \patchcmd{\titleblock@produce}
  {\frontmatter@RRAPformat}
  {\frontmatter@RRAPformat{\produce@RRAP{*#1\href{mailto:#2}{#2}}}\frontmatter@RRAPformat}
  {}{}
}%
\makeatother
\begin{document}

\preprint{AIP/123-QED}

\title{Study of stationary rigidly rotating anisotropic cylindrical fluids with new exact interior solutions of GR. IV. Radial pressure.}

\author{M.-N. C\'el\'erier}
\email{marie-noelle.celerier@obspm.fr}
\affiliation{Laboratoire Univers et Th\'eories, Observatoire de Paris, Universit\'e PSL, Universit\'e Paris Cit\'e, CNRS, F-92190 Meudon, France}

\date{\today}

\begin{abstract}
This article belongs to a series where the influence of anisotropic pressure on the gravitational properties of rigidly rotating fluids is studied using new exact solutions of GR constructed for the purpose. For mathematical simplification, stationarity and cylindrical symmetry implying three Killing vectors are considered. Moreover, two pressure components are set to vanish in turn. In Papers 1 and 2 the pressure is axially directed, while it is azimuthal in Paper 3. In present Paper 4, a radially directed pressure is considered. Since a generic differential equation, split into three parts, emerges from the field equations, three different classes of solutions can be considered. Two could only be partially integrated. The other one, that is fully integrated, yields a set of solutions with negative pressure. Physical processes where a negative pressure is encountered are depicted and give a rather solid foundation to this class of solutions. Moreover, these fully integrated solutions satisfy the axisymmetry condition while they do not verify the so-called ''regularity condition''. But, since their Kretschmann scalar does not diverge on the axis, this feature must be considered as reporting a mere coordinate singularity. Finally, the matching of these solutions to an exterior appropriate vacuum enforce other constraints on the two constant parameters defining each solution in the class. The results displayed here deserve to be interpreted in the light of those depicted in the other four papers in the series.
\end{abstract}

% insert suggested PACS numbers in braces on next line. PACS is no more maintained and will be replaced by PhySH (Physics Subject Headings), a physics classification scheme developed by APS to organize journal, meeting, and other content by topic. PhySH is currently under development.

\maketitle %\maketitle must follow title, authors, abstract and \pacs

\section{Introduction} \label{intro}

This article is the fourth in a series of five devoted to the study of anisotropic pressure in relativistic gravitating stationary fluids rigidly rotating around their axis of symmetry, through the display and analysis of new exact solutions to the field equations of General Relativity (GR). To perform this study, the pressure configuration has been specialized to three simple cases allowing total integrations of the differential equations of GR. This simplification implies that each case corresponds to the vanishing of two pressure components in the direction of the principal stresses. In papers 1 \cite{C21a} and 2 \cite{C22a}, the pressure is axially directed since its radial and azimuthal components are vanishing. In Paper 3 \cite{C21b} the pressure is azimuthally directed and its axial and radial components are zero while in present Paper 4, it is purely radial, with vanishing axial and azimuthal components. Paper 5 will be devoted to an analysis of the dust case, i.e., with no pressure, contrasted with the results obtained for the anisotropic cases, which will be themselves compared together, yielding a set of general rules.

For the axial and azimuthal pressure cases, thanks to the existence of an extra degree of freedom exhibited by the problem, general methods allowing the finding of new classes of solutions have been found and displayed in Papers 2 and 3 respectively. Now for the radial case, three classes of solutions emerge. The first one has been totally integrated without resort to any extra degree of freedom, while no integration recipe has been found for the two others. Indeed, no extra assumption has been necessary to solve the first class whose solutions are obtained directly from the field equations. And no general method has been found for the two others for which a mere simplified set of differential equations is displayed.

Taken as a whole this series of articles provide a very consistent and useful set of results allowing to initiate a new understanding of the influence of stationary rigidly rotating non perfect fluids as gravitational sources for interior spacetimes endowed with cylindrical symmetry. The choice of this symmetry has been of course driven by a mathematical simplification purpose. Three Killing vectors are involved here and contribute to the simplified behaviour of the equations to be integrated. A robust justification of the consideration of such a symmetry has been provided at length in Papers 1 and 2 to which the interested reader is referred. However, it is well known that, to begin with any study of a rather complicated issue, simplifying assumptions are welcome and are usually enough to yield significant progresses in a newly explored domain. Actually the study of anisotropic pressure in gravitating relativistic fluids is still in infancy. Moreover, even though each class of solutions considered separately can seem special, their comparison, which will be provided in Paper 5, yields new insight into the influence of the direction of the pressure.  Moreover, a number of mathematical innovations and developments have been carried out to implement the calculations, and these might be of use in other contexts.

The paper is organised as follows. In Sec. \ref{sp}, generalities about the problem and the basic equations used for solving it are provided. In Sec. \ref{solver}, three different classes of solutions are identified. In Sec. \ref{I} the metrics belonging to class I are solved and the main physical and mathematical features of the corresponding solutions are analyzed. In Sec. \ref{II}, class II fluids are considered. Section \ref{III} is devoted to the analysis of class III. The conclusions are provided in Sec. \ref{concl}.

\section{Equations describing the problem} \label{sp}

The source of the considered spacetime is a cylindrically symmetric anisotropic fluid in  stationary motion, rigidly rotating around its axis. It is dissipative and bounded by a cylindrical hypersurface $\Sigma$. Its principal stresses $P_r$, $P_z$ and $P_\phi$ satisfy the equation of state $P_z= P_{\phi}=0$. Its stress-energy tensor, whose general expression has been given by (1) of C\'el\'erier and Santos \cite{CS20}, can therefore be written under the form
\begin{equation}
T_{\alpha \beta} = (\rho + P_r) V_{\alpha}V_{\beta} + P_r g_{\alpha \beta} - P_r K_\alpha K_\beta - P_r S_\alpha S_\beta, \label{r1}
\end{equation}
where $\rho$ denotes the energy density of the fluid, $V_\alpha$, its timelike 4-velocity,  $S_\alpha$ and $K_\alpha$, spacelike 4-vectors satisfying
\begin{eqnarray}
V^\alpha V_\alpha &=& -1, \quad K^\alpha K_\alpha = S^\alpha S_\alpha = 1, \nonumber \\ 
V^\alpha K_\alpha &=& V^\alpha S_\alpha = K^\alpha S_\alpha =0. \label{r2}
\end{eqnarray}
As for the previous spacetimes of this series \cite{C21a, C22a, C21b} a spacelike hypersurface orthogonal Killing vector $\partial_z$ is assumed, in order to facilitate a subsequent junction to an exterior Lewis metric. Therefore, in geometric units $c=G=1$, the line element reads
\begin{equation}
\textrm{d}s^2=-f \textrm{d}t^2 + 2 k \textrm{d}t \textrm{d}\phi +\textrm{e}^\mu (\textrm{d}r^2 +\textrm{d}z^2) + l \textrm{d}\phi^2, \label{metric}
\end{equation}
$f$, $k$, $\mu$, and $l$ being real functions of the radial coordinate $r$ only, such as to account for stationarity. The cylindrical feature forces the coordinates to conform to the following ranges:
\begin{equation}
-\infty \leq t \leq +\infty, \quad 0 \leq r \leq +\infty, \quad -\infty \leq z \leq +\infty, \quad 0 \leq \phi \leq 2 \pi, \label{ranges}
\end{equation}
the two limits of the coordinate $\phi$ being topologically identified. These coordinates are denoted $x^0=t$, $x^1=r$, $x^2=z$, and $x^3=\phi$.

Rigid rotation allows the choice of a frame corotating with the fluid \cite{CS20,C21a,D06}. Thus, its 4-velocity can be written as
\begin{equation}
V^\alpha = v \delta^\alpha_0, \label{r3}
\end{equation}
with $v$ a function of $r$ only. Therefore, the timelike condition for $V^\alpha$ displayed in (\ref{r2}) becomes
\begin{equation}
fv^2 = 1. \label{timeliker}
\end{equation}
The first spacelike 4-vector used to define the stress-energy tensor and verifying conditions (\ref{r2}) can be chosen as
\begin{equation}
K^\alpha = - \frac{kv}{D}\delta^\alpha_0 - \frac{fv}{D}\delta^\alpha_3, \label{r4}
\end{equation}
where the $D(r)$ function is defined as
\begin{equation}
D^2 = fl + k^2. \label{D2}
\end{equation}
The second spacelike 4-vector can be written as
\begin{equation}
S^\alpha = \textrm{e}^{-\mu/2}\delta^\alpha_2.  \label{r5}
\end{equation}

\subsection{Field equations} \label{fe}

Using (\ref{r3})--(\ref{r5}) into (\ref{r1}), the components of the stress-energy tensor matching the five nonvanishing components of the Einstein tensor are obtained, and the five corresponding field equations can be written as

\begin{equation}
G_{00} = \frac{\textrm{e}^{-\mu}}{2} \left[-f\mu'' - 2f\frac{D''}{D} + f'' - f'\frac{D'}{D} + \frac{3f(f' l' + k'^2)}{2D^2}\right]= \kappa\rho f, \label{G00r}
\end{equation}
\begin{equation}
G_{03} =  \frac{\textrm{e}^{-\mu}}{2} \left[k\mu'' + 2 k \frac{D''}{D} -k'' + k'\frac{D'}{D} - \frac{3k(f' l' + k'^2)}{2D^2}\right] = - \kappa\rho k, \label{G03r}
\end{equation}
\begin{equation} 
G_{11} = \frac{\mu' D'}{2D} + \frac{f' l' + k'^2}{4D^2} = \kappa P_r \textrm{e}^{\mu}, \label{G11r}
\end{equation}
\begin{equation}
G_{22} = \frac{D''}{D} -\frac{\mu' D'}{2D} - \frac{f'l' + k'^2}{4D^2} = 0, \label{G22r}
\end{equation}
\begin{equation}
G_{33} =  \frac{\textrm{e}^{-\mu}}{2} \left[l\mu'' + 2l\frac{D''}{D} - l'' + l'\frac{D'}{D} - \frac{3l(f' l' + k'^2)}{2D^2}\right] =  \frac{\kappa \rho k^2}{f}, \label{G33r}
\end{equation}
where the primes stand for differentiation with respect to $r$.

\subsection{Conservation of the stress-energy tensor} \label{bi}

The conservation of the stress-energy tensor is implemented by the Bianchi identity, whose general form is available as (57) in C\'el\'erier and Santos \cite{CS20}. Specialized to the present case, it becomes
\begin{equation}
T^\beta_{1;\beta} = P'_r + \rho \frac{f'}{2f} + P_r\frac{D'}{D} + P_r \frac{\mu'}{2}= 0. \label{Bianchi1r}
\end{equation}
With $h(r)$ defined here as $h(r)\equiv P_r(r)/\rho(r)$, it reads
\begin{equation}
\frac{1}{2h} \frac{f'}{f} + \frac{D'}{D} + \frac{\mu'}{2} + \frac{P'_r}{P_r} = 0. \label{Bianchi2r}
\end{equation}

\subsection{Some reminders} \label{rem}

The main equations established in previous works and needed for the following calculations are briefly recalled below. The first one, initially displayed as (14) in Debbasch et al. \cite{D06}, is written here as
\begin{equation}
kf' - fk' = 2c D, \label{r6}
\end{equation}
where $c$ is an integration constant and the factor 2 is added for further convenience. This first-order ordinary differential equation in $k$ possesses, as a solution, see (24) of Paper 1,
\begin{equation}
k = f \left(c_k - 2c \int^r_0 \frac{D(v)}{f^2(v) } \textrm{d} v \right). \label{r7}
\end{equation}

Another useful relation has been established as (19) in Paper 2 and reads
\begin{equation}
\frac{f'l' +k'^2}{2D^2} = \frac{f'D'}{fD} - \frac{f'^2}{2f^2} + \frac{2c^2}{f^2}. \label{r8}
\end{equation}

\subsection{New differential equation} \label{sfe}

From the field equations and their outcome displayed in Secs. \ref{fe} and \ref{rem}, another homogeneous differential equation that will be subsequently useful can be derived. The field equation (\ref{G00r}) divided by $f$ can be written as
\begin{equation}
-\mu'' - \frac{2D''}{D} + \frac{f''}{f} - \frac{f' D'}{f D} + \frac{3(f' l' + k'^2)}{2D^2}= 2\kappa\rho \textrm{e}^{\mu}. \label{sfe1}
\end{equation}
The field equation (\ref{G11r}), with $P_r=h \rho$ inserted, becomes
\begin{equation} 
2 \kappa \rho \textrm{e}^{\mu} = \frac{1}{h}\frac{\mu' D'}{D} + \frac{1}{h}\frac{f' l' + k'^2}{2D^2}. \label{sfe2}
\end{equation}
Then, substituting (\ref{r8}) into the field equation (\ref{G22r}), one obtains
\begin{equation}
\frac{2D''}{D} = \frac{\mu' D'}{D} + \frac{f' D'}{f D} - \frac{f'^2}{2f^2} + \frac{2c^2}{f^2}. \label{sfe3}
\end{equation}
Now, (\ref{r8}), (\ref{sfe2}) and (\ref{sfe3}) are inserted into (\ref{sfe1}) which yields
\begin{eqnarray}
&-& \mu'' + \frac{f''}{f} + \frac{1-2h}{h}\left(\frac{f'^2}{2f^2} - \frac{2 c^2}{f^2}\right) - \frac{(1-h)}{h}\frac{f' D'}{f D} \nonumber \\
&-& \frac{(1+h)}{h}\frac{\mu' D'}{D} = 0. \label{sfe4}
\end{eqnarray}

This equation will be used in Sec. \ref{III} for calculations performed in the framework of Class III.

\section{Identifying three classes of solutions} \label{solver}

To allow the integration of the Bianchi identity (\ref{Bianchi2r}), a new function $F(r)$ is defined by
\begin{equation}
\frac{F'}{F} = \frac{1}{h} \frac{f'}{f}. \label{r9}
\end{equation}
Inserted into Bianchi's identity (\ref{Bianchi2r}), it gives
\begin{equation}
\frac{F'}{2F} + \frac{D'}{D} + \frac{\mu'}{2} + \frac{P'_r}{P_r}= 0, \label{r10}
\end{equation}
which can be integrated as
\begin{equation}
P_r F^{\frac{1}{2}} D \textrm{e}^{\frac{\mu}{2}} = c_B, \label{r11}
\end{equation}
where $c_B$ is an integration constant.

Now, adding (\ref{G11r}) and (\ref{G22r}), one obtains
\begin{equation}
\frac{D''}{D} = \kappa P_r \textrm{e}^{\mu}, \label{r12}
\end{equation}
where $P_r$ given by (\ref{r11}) is inserted to yield
\begin{equation}
D'' = \frac{\kappa c_B\textrm{e}^{\frac{\mu}{2}}}{F^{\frac{1}{2}}}, \label{r13}
\end{equation}
which is inserted into (\ref{G22r}), together with (\ref{r8}), to give
\begin{equation}
\left(\mu' + \frac{f'}{f} \right) D' - \left(\frac{f'^2}{2f^2} - \frac{2c^2}{f^2} \right) D - \frac{2 \kappa c_B \textrm{e}^{\frac{\mu}{2}}}{F^{\frac{1}{2}}} =0. \label{r14}
\end{equation}

Then, (\ref{G22r}) and (\ref{r12}) substituted into (\ref{G00r}), where $P_r$ is replaced by its expression given by (\ref{r11}), yield
\begin{equation}
\left(3\mu' + \frac{f'}{f} \right) D' + \left(\mu'' - \frac{f''}{f} \right) D + \frac{2 \kappa c_B \textrm{e}^{\frac{\mu}{2}}}{F^{\frac{1}{2}}} \left(\frac{1-2h}{h} \right) =0. \label{r15}
\end{equation}
Equalizing both expressions for $D'$ as given by (\ref{r14}) and (\ref{r15}), one obtains
\begin{equation}
\frac{\kappa c_B \textrm{e}^{\frac{\mu}{2}}}{F^{\frac{1}{2}}D} = - \frac{\left(\frac{f'^2}{2f^2} - \frac{2c^2}{f^2} \right) \left(3 \mu' + \frac{f'}{f} \right) + \left(\mu'' - \frac{f''}{f} \right) \left( \mu' + \frac{f'}{f} \right)}{2\left[\left(\frac{1+h}{h}\right) \mu' + \left(\frac{1-h}{h} \right) \frac{f'}{f} \right]}. \label{r16}
\end{equation}

Now, differentiating with respect to $r$ (\ref{G11r}) where (\ref{r8}) has been previously inserted and with the help of (\ref{Bianchi2r}), (\ref{r11}), (\ref{r12}), (\ref{r14}) and (\ref{r16}), then arranging, one obtains
\begin{eqnarray}
&&\left(\mu' + \frac{f'}{f} \right) \left( 2 \mu'' +  \frac{2f''}{f} - \frac{f'^2}{f^2} + \frac{4c^2}{f^2}\right) \nonumber \\
&\times& \left[\mu'' - \frac{f''}{f} + \frac{f'^2}{2f^2} - \frac{(1+h)}{2h} \frac{\mu' f'}{f} + \frac{(1-2h)}{2h} \frac{4 c^2}{f^2} \right]=0. \label{r17}
\end{eqnarray}
This equation is verified independently if any factorized expression vanishes. Each of the three corresponds to a given class of solutions which will be examined in turn below.

\section{Class I} \label{I}

\subsection{Solving for the metric functions} \label{1cmf}

Class I is composed of the solutions verifying the vanishing of the first factor in (\ref{r17}), i. e.,
\begin{equation}
\mu' + \frac{f'}{f} =0, \label{r18}
\end{equation}
which can be integrated as
\begin{equation}
f  \textrm{e}^{\mu} =c_{\mu}^2, \label{r19}
\end{equation}
where $c_{\mu}^2$ is another integration constant, squared for further purpose.

Differentiating (\ref{r19}) with respect to $r$, one obtains
\begin{equation}
\mu'' = - \frac{f''}{f} + \frac{f'^2}{f^2}. \label{r20}
\end{equation}

Inserting (\ref{r18}) and (\ref{r19}) into (\ref{r14}), one obtains
\begin{equation}
D = -  \frac{2 \kappa c_B c_{\mu}}{f^{\frac{1}{2}}F^{\frac{1}{2}} \left(\frac{f'^2}{2f^2} - \frac{2c^2}{f^2} \right) }. \label{r21}
\end{equation}  
Substituting (\ref{r19}) and (\ref{r21}) into (\ref{r11}) yields
\begin{equation}
P_r = \frac{4c^2 - f'^2}{ 4\kappa c_{\mu}^2 f}. \label{r22}
\end{equation}

The field equation (\ref{G00r}) written with $\rho = P_r/h$, and where (\ref{r8}), (\ref{r12}), (\ref{r20}) and (\ref{r22}) are inserted, becomes
\begin{equation}
\frac{D'}{D} = - \frac{f''}{f'} + \left(1- \frac{1}{4h} \right)\frac{f'}{f} + \left( \frac{1}{h} -2 \right) \frac{c^2}{f f'}. \label{r23}
\end{equation}

Then, (\ref{r22}) differentiated with respect to $r$ and divided by itself gives
\begin{equation}
\frac{P'_r}{P_r}= -  \frac{2f' f''}{4c^2 - f'^2} - \frac{ f'}{f}, \label{r24}
\end{equation}
which can be inserted into (\ref{Bianchi2r}), together with (\ref{r18}) and (\ref{r23}), such as to yield
\begin{equation}
f'' = \frac{1-2h}{4h} \left(\frac{4c^2 - f'^2}{f} \right). \label{r25}
\end{equation}

Inserting (\ref{r25}) into (\ref{r23}) gives
\begin{equation}
\frac{D'}{D} = \frac{f'}{2f}, \label{r26}
\end{equation}
which can be integrated as
\begin{equation}
D =c_D\sqrt{f}, \label{r27}
\end{equation}
where $c_D$ is another integration constant.

Then, (\ref{r18}) and  (\ref{r26}) can be substituted into (\ref{Bianchi2r}) to obtain
\begin{equation}
\frac{P'_r}{P_r}= -  \frac{1}{2h} \frac{f'}{f}. \label{r28}
\end{equation}
Now, inserting (\ref{r19}) and (\ref{r27}) into (\ref{r11}) yields
\begin{equation}
P_r F^{\frac{1}{2}}= \frac{c_B}{c_D c_{\mu}}. \label{r29}
\end{equation}

Next, (\ref{r19}) is substituted into (\ref{r13}) and gives
\begin{equation}
D'' = \frac{\kappa c_B c_{\mu}}{f^{\frac{1}{2}}F^{\frac{1}{2}}}, \label{r30}
\end{equation}
which is divided by (\ref{r21}) and yields
\begin{equation}
\frac{D''}{D} = \frac{c^2}{f^2} -\frac{f'^2}{4f^2}. \label{r31}
\end{equation}
The well-known identity,
\begin{equation}
\frac{D''}{D} = \left(\frac{D'}{D}\right)' + \frac{D'^2}{D^2}, \label{r32}
\end{equation}
with (\ref{r26}) inserted, becomes
\begin{equation}
\frac{D''}{D} = \frac{f''}{2f} - \frac{f'^2}{4f^2} \label{r33}
\end{equation}
and is then equalized to (\ref{r31}) to obtain
\begin{equation}
f'' = \frac{2c^2}{f}, \label{r34}
\end{equation}
which is inserted into (\ref{r25}) to yield
\begin{equation}
f'^2 = 4c^2 \frac{1-4h}{1-2h}, \label{r35}
\end{equation}
of which the square root reads
\begin{equation}
f' = 2 \epsilon c \sqrt{ \frac{1-4h}{1-2h}}, \label{r36}
\end{equation}
where $\epsilon=\pm 1$ determines two different subclasses of solutions.

Now, (\ref{r36}) differentiated with respect to $r$ and equalized to (\ref{r34}) gives
\begin{equation}
f = - \epsilon c \frac{(1-2h)^{\frac{3}{2}}(1-4h)^{\frac{1}{2}}}{h'}, \label{r37}
\end{equation}
which is used to divide (\ref{r36}) such as to obtain
\begin{equation}
\frac{f'}{f} = - \frac{2h'}{(1-2h)^2}, \label{r38}
\end{equation}
that can be integrated as
\begin{equation}
f = c_f^2 \exp \left(\frac{1}{2h-1} \right), \label{r39}
\end{equation}
where $c_f^2$ is an integration constant, squared for further purpose. Then, (\ref{r39}) can be inserted into (\ref{r19}), to obtain
\begin{equation}
\textrm{e}^{\mu} = \frac{c_{\mu}^2}{c_f^2} \exp \left(\frac{1}{1-2h} \right), \label{r40}
\end{equation}
which allows the coordinates $r$ and $z$ to be rescaled from a factor $c_{\mu}/c_f$, that implies $c_{\mu} = c_f$.

Similarly, (\ref{r39}) can be substituted into (\ref{r27}), that yields
\begin{equation}
D = c_D c_f \exp \left(\frac{1}{2(2h-1)} \right). \label{r41}
\end{equation}

Then, (\ref{r39}) inserted into (\ref{r37}) gives
\begin{equation}
h' = - \frac{\epsilon c}{c_f^2}(1-2h)^{\frac{3}{2}} (1-4h)^{\frac{1}{2}} \exp \left(\frac{1}{1-2h} \right), \label{r42}
\end{equation}
which it is easy to write as an equation for the radial coordinate $r$ that reads
\begin{equation}
r = - \frac{\epsilon c_f^2}{c} \int^h_{h_0} \frac{1}{(1-2v)^{\frac{3}{2}} (1-4v)^{\frac{1}{2}}} \exp \left(\frac{1}{2v-1} \right) \textrm{d} v , \label{r42b}
\end{equation}
which becomes, after integration,
\begin{equation}
r = \frac{\epsilon \sqrt{\pi}}{2  \textrm{e}^2} \frac{c_f^2}{c} \left[\textrm{erfi} \left(\sqrt{\frac{1-4h}{1-2h}}\right) - \textrm{erfi} \left(\sqrt{\frac{1-4h_0}{1-2h_0}}\right) \right], \label{r42c}
\end{equation}
where $\textrm{erfi}$ denotes the imaginary error function.

Substituting (\ref{r38}) into (\ref{r9}), one obtains
\begin{equation}
\frac{F'}{F} = - \frac{2h'}{h(1-2h)^2}, \label{r43}
\end{equation}
which can be integrated as
\begin{equation}
F = c_F^2 \frac{(1-2h)^2}{4h^2} \exp \left(\frac{2}{2h-1} \right), \label{r44}
\end{equation}
where $c_F^2$ is an integration constant, also squared for further purpose. Inserting (\ref{r44}) into (\ref{r29}), where $c_{\mu}$ has been replaced by $c_f$, yields
\begin{equation}
P_r = \frac{c_B}{c_D c_f c_F} \frac{2h}{1-2h} \exp \left(\frac{1}{1-2h} \right). \label{r45}
\end{equation}

The results displayed in (\ref{r40}), (\ref{r41}) and (\ref{r45}) can be inserted into (\ref{r12}) such as to give, after rearrangements, the following constraint on the constants that had appeared in the course of the preceding integrations. It reads
\begin{equation}
\frac{\kappa c_B}{c_D c_F}  = \frac{c^2}{ c_f^3}. \label{r47}
\end{equation}

Now, inserting (\ref{r39}), (\ref{r41}) and (\ref{r42}) into (\ref{r7}), to be able to make the usual change of integration variable at this stage, one obtains the following expression for the metric function $k$ that reads
\begin{eqnarray}
&&k = c_f^2 \exp \left(\frac{1}{2h-1}\right) \left\{c_k \right. \nonumber \\
&+& \left. \frac{2 \epsilon c_D}{c_f} \int^h_{h_0} \frac{1}{(1-2v)^{\frac{3}{2}}(1-4v)^{\frac{1}{2}}}  \exp \left(\frac{1}{2(1-2v)} \right) \textrm{d} v \right\}, \label{r48}
\end{eqnarray}
that becomes, after integration,
\begin{eqnarray}
k &=& c_f^2 \exp \left(\frac{1}{2h-1}\right) \left\{c_k - \frac{\epsilon \sqrt{2 \pi} \textrm{e}  c_D}{c_f}\left[\textrm{erf} \left(\sqrt{\frac{1-4h}{2(1-2h)}}\right) \right. \right. \nonumber \\
&-& \left. \left. \textrm{erf} \left(\sqrt{\frac{1-4h_0}{2(1-2h_0)}}\right) \right] \right\}, \label{r49}
\end{eqnarray}
where $\textrm{erf}$ denotes the error function and $h_0$, the value of $h$ at the axis where $r=0$.

Then, the function $l$ proceeds from inserting (\ref{r39}), (\ref{r41}) and (\ref{r49}) into (\ref{D2}), that gives
\begin{eqnarray}
l &=& c_D^2 -c_f^2 \exp \left(\frac{1}{2h-1}\right) \left\{c_k - \frac{\epsilon \sqrt{2 \pi} \textrm{e}  c_D}{c_f} \right. \nonumber \\
&\times& \left.  \left[\textrm{erf} \left(\sqrt{\frac{1-4h}{2(1-2h)}}\right) - \textrm{erf} \left(\sqrt{\frac{1-4h_0}{2(1-2h_0)}}\right) \right] \right\}^2, \label{r50}
\end{eqnarray}

\subsection{Behaviour of the $h(r)$ function} \label{rh}

Such as to allow physically consistent applications of these solutions, real-valuedness is demanded for the different evaluations of the functions describing them. Indeed, (\ref{r42}) can be written as
\begin{equation}
h' = - \frac{\epsilon c}{c_f^2}\sqrt{(1-2h)^3 (1-4h)} \exp \left(\frac{1}{1-2h} \right). \label{rreg9}
\end{equation}
It appears therefore that $h'$ exhibits interesting features. First, it is real-valued provided the expression under the square root is positive, which implies either $h<1/4$ or $h>1/2$. Between these limits, the derivative of the $h$ function with respect to $r$ is imaginary, which is unwanted here. Second, it keeps the same sign, i. e., that of $- \epsilon c$, everywhere inside the cylinder. Hence, the $h$ function is monotonically increasing or decreasing with the $r$ coordinate, depending on the respective signs of $\epsilon$ and $c$. Now, by convention, the parameter $c$ can be chosen positive so that $-\epsilon$ drives the monotonic behaviour of $h(r)$.

Moreover, two implicit expressions for $h(r)$ are given by (\ref{r42c}), under the form $r(h)$. The $r$ coordinate being  positive definite implies therefore: for $\epsilon = -1$, $h \geq h_0$, and for $\epsilon = +1$, $h \leq h_0$.

\subsection{Axisymmetry and regularity conditions} \label{rar}

As already recalled in Papers 1-3 \cite{C21a,C22a,C21b}, the condition to be fulfilled by an axisymmetric spacetime is that the metric function $l$ should vanish in the limit at the axis \cite{M93,P96,S09}, i.e.,
\begin{equation}
l \stackrel{0}{=} 0, \label{rax1}
\end{equation} 
where $\stackrel{0}{=}$ denotes that the relevant quantities are evaluated at the axis of symmetry where $r=0$. This implies, from (\ref{r50}), a constraint on the constants that reads
\begin{equation}
\frac{c_D^2}{c_f^2 c_k^2} =  \exp \left(\frac{1}{2h_0-1} \right), \label{rax2}
\end{equation}
which becomes, once its square root is taken,
\begin{equation}
c_D = \eta c_f c_k  \exp \left(\frac{1}{2(2h_0-1)} \right), \label{rax3}
\end{equation}
where $\eta = \pm 1$.

Now, the regularity condition corresponding to this solution reads \cite{M93,S09,C21a,C22a,C21b}
\begin{equation}
\frac{\textrm{e}^{-\mu}l'^2}{4l} \stackrel{0}{=} 1, \label{rreg1}
\end{equation}
that yields the following constraint
\begin{eqnarray}
\left\{c c_f c_k^2 \sqrt{\frac{1-4h_0}{1-2h_0}} \exp \left[\frac{1}{2(2h_0-1)} \right] - 2 \epsilon c c_D c_k \right\}^2 \nonumber \\
= c_{\mu}^2 \left[c_D^2 - c_f^2 c_k^2  \exp \left(\frac{1}{2h_0-1} \right)\right], \label{rreg2}
\end{eqnarray}
where the substitution of (\ref{rax2}) yields
\begin{equation}
h_0 = \frac{3}{4}. \label{rreg3}
\end{equation}
For this value of the ratio $h_0$, the axisymmetry constraint (\ref{rax2}) becomes
\begin{equation}
\frac{c_D^2}{c_f^2 c_k^2} =  \textrm{e}^2. \label{rax4}
\end{equation}

It is however stressed here once again that the relevance of the regularity condition is questionable. Using the Kretschmann scalar, it has been shown in Paper 2 that an axis can be no singular even though the corresponding spacetime does not fulfil the regularity condition. The apparent singularity is therefore a mere coordinate singularity. Analogous statements can be found in the literature \cite{L94,W96,C00}.

\subsection{Junction condition} \label{rjunction}

The junction conditions of a stationary rigidly rotating cylindrically symmetric source to a vacuum Weyl-Lewis spacetime have already been thoroughly worked out by Debbasch et al \cite{D06} in their Sec. 3. These authors have obtained three constraint equations: 
\begin{equation}
P_r \stackrel{\Sigma}{=} 0 \label{J1}
\end{equation}
which was already well-known from the work by Israel \cite{I66}. Examining $P_r$ as given by (\ref{r45}), it appears that this condition is indeed realized by solutions belonging to class I provided that $h_{\Sigma} = 0$.

The two other constraints are provided by Debbasch et al.'s Eqs. (37) and (38) into which their Eqs. (25) and (39) have to be inserted to obtain a relation between the parameters of the interior spacetime and those of the vacuum exterior. Adapted to the notations of the present paper, their (37) reads
\begin{equation}
2c = 2 c_{LW}, \label{J2}
\end{equation}
where the subscript LW denotes the corresponding $c$ parameter of the Lewis-Weyl metric. As it has been shown by C\'el\'erier \cite{C22a} for the axial pressure case, which can be generalized to the radial pressure configuration since the assumptions are equivalent, the $c$ parameter of the class I solutions corresponds to the value of the rotation scalar of the fluid at the axis. This yields a new interpretation of $c_{LW}$ of the Lewis-Weyl vacuum metrics, at variance as to Debbasch et al.'s claim arguing that the parameter $c_{LW}$ could be responsible, with $b$, for the stationarity of spacetime. It is indeed shown, through this rotation scalar theorem, that $c$, and therefore $c_{LW}$, is rather linked to the local rotation of the fluid source on the axis.

The last constraint exhibited by Debbasch et al., i. e., their Eq. (38), becomes, with their Eqs. (25) and (39) inserted,
\begin{equation}
1 - n + 2 b c_{LW}  \stackrel{\Sigma}{=} \frac{8 \pi}{\kappa} \frac{D f' - 2 c k'}{f}, \label{J3}
\end{equation}
where we insert (\ref{r39}), (\ref{r41}) and (\ref{r49}) so as to obtain
\begin{eqnarray}
&&1 - n + 2 b c_{LW} = \frac{8 \pi}{\kappa} \left[\frac{2 \epsilon c c_D}{c_f} \sqrt{\frac{1-4h_{\Sigma}}{1-2h_{\Sigma}}} \exp \left(\frac{1}{2(1-2h_{\Sigma})}\right) \right. \nonumber \\
&-& \left. \frac{4 \epsilon c^2 c_k}{c_f^2} \sqrt{\frac{1-4h_{\Sigma}}{1-2h_{\Sigma}}} \exp \left(\frac{1}{1-2h_{\Sigma})}\right) + \frac{4 c^2 c_D}{c_f^3} \exp \left(\frac{3}{2(1-2h_{\Sigma})}\right) \right]. \label{J4}
\end{eqnarray}
This is a rather complicated relation between the parameters of both metrics, the interior and the vacuum exterior, which does not allow a straightforward physical interpretation.

\subsection{Energy density} \label{red}

Using (\ref{r45}) into the definition of $h$ as $h= P_r/\rho$, one obtains the expression for the energy density
\begin{equation}
\rho = \frac{2 c_B}{c_D c_f c_F} \frac{1}{1-2h} \exp \left(\frac{1}{1-2h} \right). \label{red1}
\end{equation}
With (\ref{r47}) inserted, this becomes
\begin{equation}
\rho = \frac{2c^2}{\kappa c_f^4} \frac{1}{1-2h} \exp \left(\frac{1}{1-2h} \right). \label{red2}
\end{equation}

The weak energy condition, being fulfilled where $\rho\geq 0$, implies
\begin{equation}
h \leq \frac{1}{2}, \label{red3}
\end{equation}
which selects, from the results of Sec. \ref{rh}, $h<1/4$. This constraint being inconsistent with (\ref{rreg3}), the regularity condition is not fulfilled on the axis.

\subsection{Constraints from the metric signature} \label{rsign}

To be able to reproduce the proper Lorentzian signature of metric (\ref{metric}), the functions $f$, $\textrm{e}^{\mu}$, $k$ and $l$ are bounded to be all definite positive or all definite negative. Owing to the forms of (\ref{r39}) and (\ref{r40}), it is obvious that $f$ and $\textrm{e}^{\mu}$ are positive whatever the value of $h$. As regards $k$, (\ref{r49}) with (\ref{rax3}) inserted gives
\begin{eqnarray}
&&k = c_f^2 c _k \exp\left(\frac{1}{2h-1}\right)\left\{1 - \epsilon \eta \textrm{e} \sqrt{2 \pi} \exp \left(\frac{1}{2(2h_0 - 1)} \right) \right.\nonumber \\
&\times& \left.  \left[\textrm{erf} \left(\sqrt{\frac{1-4h}{2(1-2h)}}\right) - \textrm{erf} \left(\sqrt{\frac{1-4h_0}{2(1-2h_0)}}\right) \right] \right\}. \label{r51}
\end{eqnarray}
Analogously, a new expression can be obtained for $l$ by inserting (\ref{rax3}) into (\ref{r50}) which yields
\begin{eqnarray}
&&l = c_f^2 c_k^2 \left[\exp\left(\frac{1}{2h_0-1}\right) - \exp\left(\frac{1}{2h-1}\right)\left\{1 - \epsilon \eta \textrm{e} \sqrt{2 \pi}  \right. \right. \nonumber \\
&\times& \left. \left. \exp \left(\frac{1}{2(2h_0 - 1)} \right) \left[\textrm{erf} \left(\sqrt{\frac{1-4h}{2(1-2h)}}\right) \right. \right. \right.\nonumber \\
&-& \left. \left. \left. \textrm{erf} \left(\sqrt{\frac{1-4h_0}{2(1-2h_0)}}\right) \right] \right\}^2 \right]. \label{r52}
\end{eqnarray}
Since $c_f c_k$ appears as  such as a factor in $k$ and squared as a factor in $l$, one can remove them by performing a rescaling of the $\phi$ coordinate by a factor of $c_f c _k$, implying $c_f c_k = 1$ such as to preserve the range $0\leq \phi \leq 2\pi$. Therefore, these metric functions become
\begin{eqnarray}
&& k = c_f \exp\left(\frac{1}{2h-1}\right)\left\{1 - \epsilon \eta \textrm{e} \sqrt{2 \pi} \exp \left(\frac{1}{2(2h_0 - 1)} \right) \right. \nonumber \\
&\times& \left. \left[\textrm{erf} \left(\sqrt{\frac{1-4h}{2(1-2h)}}\right) - \textrm{erf} \left(\sqrt{\frac{1-4h_0}{2(1-2h_0)}}\right) \right] \right\}, \label{r53}
\end{eqnarray}
\begin{eqnarray}
l &=& \exp\left(\frac{1}{2h_0-1}\right) - \exp\left(\frac{1}{2h-1}\right)\left\{1 - \epsilon \eta \textrm{e} \sqrt{2 \pi}  \right.  \nonumber \\
&\times& \left. \exp \left(\frac{1}{2(2h_0 - 1)} \right) \left[\textrm{erf} \left(\sqrt{\frac{1-4h}{2(1-2h)}}\right)  \right. \right. \nonumber \\
&-& \left. \left. \textrm{erf} \left(\sqrt{\frac{1-4h_0}{2(1-2h_0)}}\right) \right] \right\}^2. \label{r54}
\end{eqnarray}
The following constraints are therefore imposed on the solutions. From the positiveness of $k$ and owing to that, assumed, of $c_f$ and to that, intrinsic, of the exponential function, it reads
\begin{eqnarray}
&& 0 \leq 1 - \epsilon \eta \textrm{e} \sqrt{2 \pi} \exp \left(\frac{1}{2(2h_0 - 1)}\right) \left[\textrm{erf} \left(\sqrt{\frac{1-4h}{2(1-2h)}}\right) \right. \nonumber \\
&-& \left.  \textrm{erf} \left(\sqrt{\frac{1-4h_0}{2(1-2h_0)}}\right) \right]. \label{r55}
\end{eqnarray}
And from the positiveness of $l$, one obtains
\begin{eqnarray}
\exp\left(\frac{1}{2h_0-1}- \frac{1}{2h-1}\right) \geq \left\{1 - \epsilon \eta \textrm{e} \sqrt{2 \pi} \exp \left(\frac{1}{2(2h_0 - 1)} \right) \right. \nonumber \\
\times \left. \left[\textrm{erf} \left(\sqrt{\frac{1-4h}{2(1-2h)}}\right) - \textrm{erf} \left(\sqrt{\frac{1-4h_0}{2(1-2h_0)}}\right) \right] \right\}^2. \label{r56}
\end{eqnarray}
Both inequalities (\ref{r55}) and (\ref{r56}) set new limits on the allowed intervals for $h$, depending on the actual value of $h_0$, provided it is compatible with $h_0<1/4$, and depending also on the respective signs of $\epsilon$ and $\eta$.

\subsection{Subclass Ia: class I with $\epsilon = +1$} \label{Ia}

This subclass comprises the solutions obtained for $\epsilon = +1$. With this value of $\epsilon$, the constraints displayed above give:

- from the positiveness of the $r$ coordinate:
\begin{equation}
h \leq h_0. \label{Ib1}
\end{equation}

- the derivative $h'$, being negative in this case, implies, when coupled to the junction condition that the $h(r)$ function is decreasing with $r$ from $h_0$ to $h_{\Sigma}$ and that
\begin{equation}
h_{\Sigma} = 0 \leq h \leq h_0 \leq \frac{1}{4}, \label{Ib2}
\end{equation}
which imposes $h \geq 0$ on the whole spacetime. Therefore, imposing the weak energy condition involves satisfying the strong one, i.e., $P_r \geq 0$. 

- the status of the solutions as regards the positiveness of the metric functions $k$ and $l$ depends also of the sign of $\eta$.
The inequality (\ref{r55}) issued from $k>0$ implies that $\eta = -1$ whatever $h_0 > 0$, while the constraint (\ref{r56}) corresponding to $l>0$ is never fulfilled whatever the sign of $\eta$ and the positive value of $h_0$. Hence, the metric signature is improper and the subclass Ib is ruled out.

\subsection{Subclass Ib: class I with $\epsilon = -1$} \label{Ib}

This subclass includes all the class I solutions obtained with $\epsilon = -1$. This value of $\epsilon$ inserted into the constraints displayed above gives:

- from the positiveness of the $r$ coordinate:
\begin{equation}
h_0 \leq h. \label{Ia1}
\end{equation}

- from the derivative $h'$ coupled to the junction condition: $h'$ being positive in this case, the $h(r)$ function is increasing with $r$ from $h_0$ to $h_{\Sigma}$ that gives
\begin{equation}
h_0 \leq h \leq h_{\Sigma} < \frac{1}{4} , \label{Ia2}
\end{equation}
which is indeed compatible with the junction condition
\begin{equation}
h_{\Sigma} = 0, \label{Ia3}
\end{equation}
and therefore
\begin{equation}
h_0 \leq h \leq h_{\Sigma} = 0, \label{Ia4}
\end{equation}
which imposes $h \leq 0$ on the whole spacetime. Coupled to the weak energy condition, this yields $P_r \leq 0$. 

- as regards the inequalities enforced on the solutions by the positiveness of the metric functions $k$ and $l$, the constraints depend not only on the sign of $\epsilon$, but also on that of $\eta$. 

A straightforward analysis of inequality (\ref{r55}), induced by $k>0$, shows that, with $\eta = -1$, the metric function $k$ is positive definite for any negative value of $h_0$, while for $\eta = +1$, $k$ is positive definite for $h_0$ values verifying $h_0 \leq h_1$, where $h_1$ is defined by
\begin{equation}
1 = \textrm{e} \sqrt{2 \pi} \exp \left(\frac{1}{2(2h_0 - 1)}\right) \left[\textrm{erf} \left(\sqrt{\frac{1-4h_1}{2(1-2h_1)}}\right) - \textrm{erf} \sqrt{\frac{1}{2}} \right]. \label{Ia5}
\end{equation}

Now, the study of the inequality (\ref{r56}), linked to the condition $l>0$, yields two constraints. The first demands $\eta = +1$. The second imposes that $h_0 > h_2$, with $h_2$ defined by
\begin{equation}
\frac{1-4h_2}{1-2h_2} = \left[ 1 + \exp \left(\frac{1}{2(1-2h_2)} - 1 \right) \right]^2. \label{Ia6}
\end{equation}

Hence, the summary of these constraints induced by the metric signature is first that, $\eta = +1$, and second that $h_0$ must be larger than the highest among $h_1$ and $h_2$. Provided such 
conditions are satisfied, the solutions of this subclass are properly displayed and characterized.

\subsection{Singularities} \label{rsing}

Owing to their above expressions, the four metric functions, the energy density and the radial pressure are all diverging for $h$ or $h_0$ equal to $1/2$, which is situated outside the interval allowed to this ratio.

Now, the functions $k$ and $l$ vanish on the axis, but neither the energy density nor the pressure diverge or vanish there. Moreover the Krestschmann scalar, $K$, computed using the SageMath \cite{S22} and Mathematica \cite{M22} softwares, reads
\begin{equation}
K = 16 c^4 \exp\left(\frac{2}{1-2h} - \frac{2}{1-2h_0}\right) \frac{(3-13h+19h^2)}{(1-2h)^2}. \label{sing1}
\end{equation}
The only value of $h$ for which $K$ diverges is indeed the unreachable $h=1/2$, while, for $h=h_0$, $K$ exhibits a finite value. Therefore, the axis must be considered as a mere coordinate singularity. Indeed, the regularity condition is not satisfied at this locus. However, a discussion about the relevance of this notion has been given in Paper 2 and its conclusions can be applied here to explain out any drawback purportedly linked to such a feature.

\subsection{Hydrodynamical scalars, vectors and tensors} \label{rhydro}

In this section, the expressions for the acceleration vector, the rotation or twist tensor and the shear tensor of the fluid are displayed \cite{CS20,C21a,C22a}. The nonzero component of the acceleration vector being
\begin{equation}
\dot{V}_1 = \frac{1}{2} \frac{f'}{f}, \label{rdotV1a}
\end{equation}
it becomes, with (\ref{r38}) and (\ref{r42}) inserted,
\begin{equation}
\dot{V}_1 = \frac{\epsilon c}{c_f^2} \sqrt{\frac{1-4h}{1-2h}}  \exp \left(\frac{1}{1-2h} \right). \label{rdotV1b}
\end{equation}
Then, from, e. g., (82) of Paper 1, the modulus of the acceleration vector can be written
\begin{equation}
\dot{V}^\alpha \dot{V}_\alpha = \frac{1}{4} \frac{f'^2}{f^2} \textrm{e}^{-\mu}, \label{rmodaccela}
\end{equation}
where (\ref{r38}), (\ref{r40}) and (\ref{r42}) are inserted to yield
\begin{equation}
\dot{V}^\alpha \dot{V}_\alpha = \frac{c^2}{c_f^4} \frac{1-4h}{1-2h} \exp \left(\frac{1}{1-2h} \right). \label{rdotV1c}
\end{equation}

Now, the nonvanishing component of the rotation tensor follows from, e. g., (81) of Paper 1, as
\begin{equation}
2 \omega_{13} = - (2kv'+k'v), \label{romega13a}
\end{equation}
which becomes, owing to the use of (115) of C\'el\'erier and Santos \cite{CS20} and of (\ref{r6}),
\begin{equation}
2 \omega_{13} = \frac{2 c D}{f^{\frac{3}{2}}}, \label{romega13b}
\end{equation}
where inserting (\ref{r39}) and (\ref{r41}) yields
\begin{equation}
2 \omega_{13} = 2  c_f^4 \exp \left(\frac{2h+1}{2h-1} \right). \label{romega13c}
\end{equation}
Then, the rotation scalar follows from (86) of Paper 1 as
\begin{equation}
\omega^2 = \frac{c^2}{f^2\textrm{e}^{\mu}}, \label{romega2a}
\end{equation}
which becomes, after inserting (\ref{r39}) and (\ref{r40}),
\begin{equation}
\omega^2 = \frac{c^2}{c_f^4} \exp \left(\frac{1}{1-2h} \right). \label{romega2b}
\end{equation}

Here, the proposition demonstrated in Appendix A of Paper 2 can be used to derive a new constraint on the constant parameters of this class. It states indeed that any class of solutions of the type studied here, verifying (\ref{r6}) together with an equation of the form of (A1) of Paper 2, which corresponds here to (\ref{r18}), satisfies
\begin{equation}
\omega^2 \stackrel{0}{=} c^2, \label{romega2c}
\end{equation}
which, inserted into (\ref{romega2b}) written at the axis, yields
\begin{equation}
c_f^4 = \exp \left(\frac{1}{1-2h_0}\right). \label{romega2d}
\end{equation}

As already well-known, the shear tensor vanishes for any rigidly rotating fluid \cite{D06,C21a}.

\subsection{Behaviour of the energy density and of the pressure} \label{bep}

Now the behaviour of the energy density and of the radial pressure can be analyzed. The constraint (\ref{red3}) on $h$ from the weak energy condition is indeed consistent with (\ref{Ia4}) from the junction condition coupled to the behaviour of $h(r)$, which imply $h_{\Sigma} = 0$.

Differentiating the expression for $\rho$, as given by (\ref{red2}), with respect to $h$, one obtains
\begin{equation}
\frac{\textrm{d}\rho}{\textrm{d}h} = \frac{8c^2}{\kappa c_f^4} \frac{1-h}{(1-2h)^3} \exp \left(\frac{1}{1-2h} \right). \label{red4}
\end{equation}
Since $h<1/4$, the above expression is positive definite which implies that, for the only well-behaved subclass Ib, $\rho$ is monotonically increasing with $h$. Moreover, is has been shown in Sec. \ref{Ib} that $h$ is monotonically increasing with $r$. Therefore, the energy density $\rho$ is also monotonically increasing from the axis to the $\Sigma$ boundary.

A straightforward more thorough analysis, using (\ref{romega2d}) into (\ref{red2}), shows that the function $\rho(h)$ is indeed increasing from
\begin{equation}
\rho(h_0) = \frac{2c^2}{\kappa} \frac{1}{1-2h_0}\label{red5}
\end{equation}
to
\begin{equation}
\rho(h_{\Sigma}=0) = \frac{2c^2}{\kappa} \exp\left(\frac{-2h_0}{1-2h_0} \right). \label{red6}
\end{equation}

Since $h(r)$ increases from $h_0<0$ to $h_{\Sigma}=0$, the pressure $P_r= h \rho$ monotonically increases from $P_r(h_0)<0$ to $P_r(h_{\Sigma})=0$.

This case is therefore fully characterized and consistent for a fluid with a negative pressure.

\subsection{Class I summary} \label{Isum}

The main properties of class I, which has been defined as the set of solutions satisfying (\ref{r18}), an equation directly issued from the field equations themselves without need of any added assumption on the form of the metric functions, can be summarized as follows. Its metric functions, with all the constraints displayed in Sections \ref{rh}--\ref{bep} inserted, read
\begin{equation}
f = \exp \left[\frac{1}{2(1-2h_0)} + \frac{1}{2h-1} \right], \label{sum1}
\end{equation}
\begin{equation}
\textrm{e}^{\mu} = \exp \left(\frac{1}{1-2h} \right), \label{sum2}
\end{equation}
\begin{eqnarray}
&& k = \exp\left[\frac{1}{4(1-2h_0)} +\frac{1}{2h-1}\right]\left\{1 + \textrm{e} \sqrt{2 \pi} \exp \left(\frac{1}{2(2h_0 - 1)} \right) \right. \nonumber \\
&\times& \left. \left[\textrm{erf} \left(\sqrt{\frac{1-4h}{2(1-2h)}}\right) - \textrm{erf} \left(\sqrt{\frac{1-4h_0}{2(1-2h_0)}}\right) \right] \right\}, \label{sum3}
\end{eqnarray}
\begin{eqnarray}
l &=& \exp\left(\frac{1}{2h_0-1}\right) - \exp\left(\frac{1}{2h-1}\right)\left\{1 + \textrm{e} \sqrt{2 \pi}  \right.  \nonumber \\
&\times& \left. \exp \left(\frac{1}{2(2h_0 - 1)} \right) \left[\textrm{erf} \left(\sqrt{\frac{1-4h}{2(1-2h)}}\right)  \right. \right. \nonumber \\
&-& \left. \left. \textrm{erf} \left(\sqrt{\frac{1-4h_0}{2(1-2h_0)}}\right) \right] \right\}^2. \label{sum4}
\end{eqnarray}

The behaviour of the $h(r)$ function has been analyzed from
\begin{equation}
h' = c \exp \left[\frac{1}{2(2h_0 - 1)} + \frac{1}{1-2h} \right] (1-2h)^{\frac{3}{2}} (1-4h)^{\frac{1}{2}} , \label{sum5}
\end{equation}
\begin{eqnarray}
r &=& \frac{\sqrt{\pi}}{2 \textrm{e}^2 c} \exp\left[\frac{1}{2(1-2h_0)}\right] \left[\textrm{erfi} \left(\sqrt{\frac{1-4h_0}{1-2h_0}}\right) \right. \nonumber\\
&-& \left. \textrm{erfi} \left(\sqrt{\frac{1-4h}{1-2h}}\right) \right]. \label{sum6}
\end{eqnarray}

The energy density and the pressure read
\begin{equation}
\rho = \frac{2c^2}{\kappa} \frac{1}{1-2h} \exp \left(\frac{1}{1-2h} - \frac{1}{1-2h_0}\right), \label{sum7}
\end{equation}
\begin{equation}
P_r = \frac{2c^2}{\kappa} \frac{h}{1-2h} \exp \left(\frac{1}{1-2h} - \frac{1}{1-2h_0}\right). \label{sum8}
\end{equation}

The two remaining auxiliary functions are given here for completeness as
\begin{equation}
D = \exp \left[\frac{1}{2(2h-1)} + \frac{1}{4(2h_0 - 1)}\right], \label{sum9}
\end{equation}
\begin{equation}
F = \frac{\kappa^2 c_B^2}{c^4 c_D^2} \frac{(1-2h)^2}{4h^2} \exp \left[\frac{2}{2h-1} + \frac{3}{2(1-2h_0)}\right]. \label{sum10}
\end{equation}

Moreover, the solutions of this class exhibit the following properties:

- they verify the weak energy condition but not the strong one which means that the pressure $P_r$ is negative everywhere. Hence, the use of such solutions must be reserved for the study of fluids exhibiting negative pressure as gravitational sources. 

- the ratio $h$ between the pressure and the energy density increases from $h_0$ on the axis up to $h_{\Sigma}$ at the boundary where the vanishing of the pressure ensure a proper matching to the exterior Lewis-Weil metric. The bottom value $h_0$ is itself lower bounded by the largest among the two limiting ratios $h_1$ defined by (\ref{Ia5}) or $h_2$ defined by (\ref{Ia6}).

- these solutions verify the axisymmetry condition, not the regularity condition but, since the Kretschmann scalar does not diverge at the axis, this locus must be considered as a coordinate singularity.

\section{Class II} \label{II}

This class is composed of the solutions verifying the vanishing of the second factor in (\ref{r17}), i. e.,
\begin{equation}
\mu'' +  \frac{f''}{f} - \frac{f'^2}{2f^2} + \frac{2c^2}{f^2} =0. \label{II1}
\end{equation}

With (\ref{r13}) inserted, (\ref{r14}) gives
\begin{equation}
\mu' = \left(\frac{f'^2}{2f^2} - \frac{2c^2}{f^2}\right) \frac{D}{D'} + \frac{2D''}{D'} - \frac{f'}{f}. \label{II2}
\end{equation}
Then, (\ref{r13}), (\ref{II1}) and (\ref{II2}) are substituted into (\ref{r15}) such as to obtain
\begin{equation}
\left(\frac{1+h}{h}\right)\frac{D''}{D} - \frac{f'D'}{fD} -  \frac{f''}{f} + \frac{f'^2}{f^2} - \frac{4c^2}{f^2} =0. \label{II3}
\end{equation}
Now, (\ref{r14}) with (\ref{r13}) inserted can be written as
\begin{equation}
\frac{f'^2}{f^2} - \frac{4c^2}{f^2} = -\frac{4D''}{D} +  2\left( \mu' + \frac{f'}{f}\right) \frac{D'}{D}, \label{II4}
\end{equation}
which can be substituted into (\ref{II3}) and yields
\begin{equation}
\frac{f''}{f} - \frac{f'D'}{fD} = \left(\frac{1-3h}{h}\right)\frac{D''}{D} +  2\mu' \frac{D'}{D}, \label{II5}
\end{equation}
which, once inserted into (\ref{r15}), together with (\ref{r13}), gives an equation implying only $D$ plus its derivatives and the derivatives of $\mu$. This equation reads
\begin{equation}
\mu''  + \frac{D'}{D} \mu' + \left(\frac{1-h}{h}\right)\frac{D''}{D} = 0. \label{II6}
\end{equation}
It can be viewed as a first order differential equation in $\mu'$ whose general solution is
\begin{equation}
\mu' = \frac{1}{D} \left[c_{\mu} - \int^r_0 \left(\frac{1-h(r)}{h(r)}\right) D''(r) \textrm{d}r\right]. \label{II6}
\end{equation}

As it has already been stressed in Paper 2, up to now, only five initial independent differential equations have been made available for six unknowns, i.e., the four metric functions, $f$, $k$, $\textrm{e}^{\mu}$, $l$, the energy density $\rho$ and the pressure $P_z$, alternatively the ratio $h$. Therefore, the set of equations needs to be closed by an additional constraint which, in Papers 1 and 2, amounted to the expression for the metric function $f(h)$. At the end of the calculations, this constraint would transform into properties of the physical features of the fluid, which can thus be selected for further applications.

Here, the simplifications made to the original field equations have reduced the mathematical problem to the resolution of only three equations to be solved is at this stage:

- (\ref{II1}), involving only $\mu''$ plus $f$ and derivatives.

- (\ref{II3}), implying only $f$, $D$ and their derivatives.

- (\ref{II6}), implying only $D$, $D''$ and $\mu'$.

They stand for four unknowns, $f$, $\mu$, $D$ and $h$ (alternatively $h'$). Their thorough examination leads to strongly suspect that no assumption designed to make use of the extra degree of freedom can lead to the finding of an analytical solution for the metric functions and the associated physical properties of the source. However, since this triplet represents a simplification of the field equation, it is here displayed as such.

\section{Class III} \label{III}

This class is composed of the solutions verifying the vanishing of the third factor in (\ref{r17}), i. e.,
\begin{equation}
\mu'' -  \frac{f''}{f} - \frac{(1+h)}{2h}\frac{\mu' f'}{f} + \frac{f'^2}{2f^2} + \frac{(1-2h)}{2h}\frac{4c^2}{f^2} =0. \label{III1}
\end{equation}
Since (\ref{II2}) still applies in this case, it can be inserted into (\ref{r15}) together with (\ref{r13}) such as to obtain
\begin{equation}
\mu'' =  \frac{f''}{f} - 3 \left(\frac{f'^2}{2f^2} - \frac{2c^2}{f^2} \right) + \frac{2(1+h)}{h}\frac{D''}{D} -  \frac{2f'D'}{fD}. \label{III2}
\end{equation}
Then, the substitution of (\ref{II2}) and (\ref{III2}) into (\ref{III1}) yields
\begin{eqnarray}
&& \left(\frac{1+h}{h}\right) \left(\frac{2}{D} - \frac{f'}{fD'}\right)D'' - \left(\frac{1+h}{2h}\right)\left(\frac{f'^2}{2f^2} - \frac{2c^2}{f^2} \right)\frac{f'D}{fD'} \nonumber \\
&-& \frac{2f'D'}{fD} + \left(\frac{1-h}{2h}\right) \frac{f'^2}{f^2} + \left(\frac{1+h}{h}\right) \frac{2c^2}{f^2} = 0. \label{III3}
\end{eqnarray}
Now, the Class III defining equation (\ref{III1}) can be written as
\begin{eqnarray}
&-&\mu'' +  \frac{f''}{f} + \frac{1-2h}{h} \left(\frac{f'^2}{2f^2} - \frac{2c^2}{f^2}\right) = - \frac{(1+h)}{2h}\frac{\mu' f'}{f} \nonumber \\
&+& \frac{(1-h)}{h}\frac{f'^2}{2f^2}, \label{III4} 
\end{eqnarray}
which, inserted into (\ref{sfe4}), gives
\begin{equation}
(1-h)\frac{f'}{f} \left(\frac{f'}{2f} - \frac{D'}{D}\right) = (1+h)\mu' \left(\frac{f'}{2f} + \frac{D'}{D}\right). \label{III5}
\end{equation}

Even though two equations, (\ref{III1}) and (\ref{III3}), where the metric functions appear two by two, and a third one which is first order in three of them, have been obtained, the issue appears more involved than in the case of Class II. The study of both these classes is therefore postponed to future work and only class I will be further considered for the moment.

\section{Conclusions} \label{concl}

The third configuration closing the present study of interior spacetimes gravitationally sourced by a stationary rigidly rotating nonperfect fluid, using new exact solutions of Einstein's field equations, has been considered. This set, where the anisotropic pressure is radially directed, includes three classes, generated each by a differential equation issued from a factorization of the field ones. The mathematically simplest among these three classes, denoted Class I, has been fully integrated without resorting to any extra assumption, which could mean that the choice of the equation defining this class is equivalent, as regards the number of degrees of freedom, to the choice of the expression of a given metric function as it was done previously to obtain the solutions in Papers 1-3. For the other two classes, Class II and Class III, some partially integrated equations have been provided, but they could not be fully integrated as it has been successfully done for Class I.

Solutions of the latter have been properly matched to an exterior Lewis vacuum such as to constitute a potential candidate physically well-behaved. Doing so, the interpretation of the Lewis-Weyl vacuum parameter $c$, denoted here $c_{LW}$, has been corrected. At variance as to Debbasch et al.'s claim arguing that the parameter $c_{LW}$ could be responsible, with $b$, for the stationarity of spacetime \cite{D06}, it has been shown here that $c_{LW}$ is rather linked to the local rotation of the fluid source on the axis.  

Moreover, while they verify the axisymmetry condition, class I solutions do not satisfy the regularity condition on the axis. In fact, it has been shown that the realization of a precise value of the ratio $h$ on the axis, would allow this regularity condition to be fulfilled. However, this particular value of $h_0$ is excluded from the validity interval of this ratio issuing from the weak energy constraint. Now, since the Kretschmann scalar does not diverge at the axis, this locus must be considered as a mere coordinate singularity. This feature confirms once again that the ''regularity condition'', if it is indeed a sufficient condition for a regular axis is by no mean a necessary one \cite{W96,C00}.

Moreover, while they satisfy the weak energy condition, these spacetimes do not verify the strong one, which implies that the pressure is negative. Hence, such solutions must be devoted to the study of fluids exhibiting negative pressure as gravitational sources. Now, negative pressure states for ''standard'' liquids are metastable  with respect to the liquid-vapour phase transition. In laboratory, this metastable state is reached by stretching and the liquids can withstand negative pressure for days. In nature, the phenomenon of cavitation is an example of such metastable states \cite{M02}. The issue whether metastable states can be exhibited by some astrophysical objects is therefore open, while it is out of the scope of the present work. Moreover, negative pressure has also been invoked in cosmology to explain ''dark energy''. Hence, there are indeed physically robust cases when a negative pressure fluid can occur.

Other interesting applications might tentatively involve the framework of cosmic (super)strings. Indeed an infinitely extended cylinder could be considered as a valid approximation for such strings whose width is indeed so small that they are generally studied in the zero-width approximation. Their internal structure is therefore neglected and a method to analyze it might be of use. If it should appear that a negative pressure could apply in some appropriate physical cases, Class I solutions might close the set, initiated in Papers 1 to 3, of spacetimes exhibiting three complementary anisotropic negative pressures and likely to provide a first insight into the issue of anisotropic pressure in GR.

The tentative resolution of Class II and Class III have led to reduce both problems to the integration of three sets of differential equations involving four unknowns, i. e., two metric functions, $f$ and $\mu$, and two auxiliary functions, $D$ and $h$. Once these equations potentially solved for these four functions, the two other metric functions arise as in Papers 1-3. However, contrary to what has been successfully completed for Class I and in Papers 1-3, it has not been possible to find here a closure constraint equation able to lead to fully integrated analytic solutions. Therefore, the available exact solutions for the radial pressure case are currently restricted to the negative pressure Class I.

A number of remarks and discussions can issue from the results obtained in this set of four papers devoted to the analysis of anisotropic pressure with new exact analytical solutions of GR. They will be displayed in a last companion paper, Paper 5, where, in particular, the dust limit and the regularity condition will be examined.

Moreover, in this article, and in the four companion papers, the fluid has been assumed to rotate rigidly. Works considering an analogous issue, but applied to differentially rotating fluids, are currently in progress. Such a task, when completed, might add further insights into some features of gravitation theory, as the current series of papers have done.

\acknowledgments

The author wants to acknowledge the precious help of Renaud Savalle for running the SageMath and Mathematica softwares used for the calculation of the Kretschmann scalar.

\end{document}